\documentclass[aps,prx,onecolumn,superscriptaddress]{revtex4-2}
\usepackage{amssymb}
\usepackage{amsmath,bm}
\usepackage{graphicx,color}
\usepackage{bbm}
\usepackage{newtxmath}
\usepackage{multirow}
\usepackage{xcolor}
\usepackage{hyperref}
\usepackage{subcaption}
\usepackage{caption}

\usepackage{pgfplots}
\pgfplotsset{width=\columnwidth,compat=1.9}

\newcommand{\be}{\begin{equation}}
\newcommand{\ee}{\end{equation}}
\newcommand{\bea}{\begin{eqnarray}}
\newcommand{\eea}{\end{eqnarray}}

\newcommand{\A}{\mathcal{A}}
\newcommand{\g}{\mathbf{g}}

\newcommand{\gbar}{{\mathbf{g}^{0}}}
\newcommand{\Gammabar}{{\Gamma^{0}}}
\newcommand{\Deltabar}{\nabla^{2}_{\gbar}}

\newcommand{\figref}[1]{\figurename~\ref{#1}}
\newcommand{\appref}[1]{App.~(\ref{#1})}
\renewcommand{\eqref}[1]{Eq.~(\ref{#1})}

\newcommand{\chione}{\chi^{(1)}}
\newcommand{\chitwo}{\chi^{(2)}}
\newcommand{\phione}{\phi^{(1)}}
\newcommand{\phitwo}{\phi^{(2)}}
\newcommand{\psione}{\psi^{(1)}}

\newcommand{\etatwo}{\eta^{(2)}}
\newcommand{\KM}{Kolosov-Muskhelishvili }

\begin{document}

\title{Nonlinear extension of \KM stress function formalism}

\author{Oran Szachter}
\affiliation{Racah Institute of Physics, The Hebrew University of Jerusalem, Jerusalem, Israel 9190401}
\author{Eytan Katzav}
\affiliation{Racah Institute of Physics, The Hebrew University of Jerusalem, Jerusalem, Israel 9190401}
\author{Mokhtar Adda-Bedia}
\affiliation{Université de Lyon, Ecole Normale Supérieure de Lyon, CNRS, Laboratoire de Physique, Lyon F-69342, France}
\author{Michael Moshe}
\email{michael.moshe@mail.huji.ac.il}
\affiliation{Racah Institute of Physics, The Hebrew University of Jerusalem, Jerusalem, Israel 9190401}


\begin{abstract}
The method of stress-function in elasticity theory is a powerful analytical tool with applications to a wide range of physical systems, including defective crystals, fluctuating membranes, and more. A complex coordinates formulation of stress function, known as \KM formalism, enabled the analysis of elastic problems with singular domains, particularly cracks, forming the basis for fracture mechanics.   
A shortcoming of this method is its limitation to linear elasticity, which assumes Hookean energy and linear strain measure. 
Under finite loads, the linearized strain fails to describe the deformation field adequately, reflecting the onset of geometric nonlinearity. The latter is common in materials experiencing large rotations, such as regions close to the crack tip or elastic metamaterials. 
While a nonlinear stress function formalism exists, \KM complex representation had not been generalized and remained limited to linear elasticity. This paper develops a \KM formalism for nonlinear stress function. The new formalism allows us to port methods from complex analysis to nonlinear elasticity and to solve nonlinear problems in singular domains.
Upon implementing the method to the crack problem, we discover that nonlinear solutions strongly depend on the applied remote loads, excluding a universal form of the solution close to the crack tip and questioning the validity of previous studies of nonlinear crack analysis.

\end{abstract}

\maketitle

\section{Introduction}

Elasticity theory is the main pillar in studying and analyzing complex physical phenomena in solids, including defects mediated plasticity in crystals \cite{seung1988defects}, fracture \cite{broberg1999cracks}, wrinkling, and growth of living matter \cite{lidmar2003virus,armon2011geometry,mirabet2011role}. A central motif in these examples is the need to find the stressed state of a system prior to a more detailed analysis that accounts for additional mechanisms. Under the requirement of mechanical equilibrium, finding the stress field $\boldsymbol{\sigma}$ reduces to solving a bulk force balance constitutive equation
\begin{equation}
    \mathrm{div} \, \boldsymbol{\sigma} = 0\;,
    \label{eq:Equilibrium}
\end{equation}
accompanied by conditions on the traction forces or displacement on the boundaries \cite{landau1959course}.
For example, in the case of crack propagation, upon solving \eqref{eq:Equilibrium}, balancing relaxation of elastic energy with fracture energy at the vicinity of the crack tip determines crack trajectory \cite{broberg1999cracks}.

One particularly useful method to solve \eqref{eq:Equilibrium} in 2D elasticity is by representing the stress in terms of a single scalar elastic potential $\chi$, formally denoted as \cite{airy1862strains, Kupferman2021doubleI,Kupferman2021doubleII}
\begin{equation}
    \boldsymbol{\sigma} = \mathrm{curl}\,\mathrm{curl} \chi.
\end{equation}
The potential $\chi$, also known as the Airy stress function, is determined by a geometric compatibility equation. Within linear elasticity, this condition reduces to a differential relation of the strain tensor, leading to the famous biharmonic equation \cite{landau1959course}
\begin{equation}
    \nabla^2\nabla^2 \chi = 0 \;.
    \label{eq:Rep}
\end{equation}
Airy's stress function approach is one of the main tools for solving problems in linear elasticity, with a wide range of applications, including statistical physics of fluctuating membranes \cite{Nelson2002defects,Kosmrlj2013mechanical}, assemblies of structural defects as screening elastic fields \cite{Bausch2003grain}, wrinkling patterns in thin sheets \cite{Tobasco2021curvature}, and granular matter \cite{Henkes2009statistical}.

For highly symmetric domains and loadings, the biharmonic equation is analytically solvable. In contrast, in problems with singular boundaries, e.g., the crack problem, or problems with low symmetry, solving the biharmonic equation is a daunting task. An elegant and advantageous mathematical method for analyzing \eqref{eq:Rep} is the use of complex coordinates, known as
the Kolosov-Muskhelishvili formalism \cite{muskhelishvili1946singular,muskhelishvili_problems,broberg1999cracks}. 
For example, the crack problem is canonically solved within this formalism, resulting in the celebrated universal $1/\sqrt{r}$ stress singularity at the crack tip, and forming the basis for Linear Elastic Fracture Mechanics \cite{broberg1999cracks}. 

Note that the stress function formalism, particularly the Kolosov-Muskhelishvili formalism, is limited to linear elasticity, which assumes {\it two} distinct linearizations: (i) Hookean elasticity with small strains and (ii) strain that is linear in deformation gradients. 
However, in the case of fracture, the elastic solution for the stresses diverges at the crack tip, where the fracture process occurs, and this questions the validity of the linear elastic solution in this region.

Indeed, experiments directly identified deviations from linear elasticity close to the crack tip and therefore called for nonlinear analysis of the elastic crack problem \cite{livne2008breakdown}.  Nonlinearity, in this case, is rooted in the appearance of large rotations, for which a nonlinear measure of strain is needed, and is geometric in nature. Furthermore, a series of publications reported nonlinear perturbative analysis of the asymptotic solution close to the crack tip \cite{bouchbinder2008weakly,bouchbinder2010weakly,livne2008breakdown}.
Since asymptotic analysis lacks the necessary and sufficient boundary conditions (e.g., stress at infinity), a new type of boundary condition was invoked to select the nonlinear solution \cite{bouchbinder2010weakly}. This condition, the requirement of divergence-free stress on the crack tip, determines a stress singularity of  $~1/r$, stronger than the singularity predicted by the linear theory. We claim that the need for additional boundary conditions reflects the absence of a systematic methodology for solving the complete problem that accounts for the remote boundary conditions. 

In this paper, we build on a previously suggested nonlinear generalization of Airy's stress function approach \cite{moshe2014plane,moshe2015elastic,Bar-Sinai2020} and reformulate it using a complex function representation. Thus, our method forms a nonlinear generalization of the \KM formalism, allowing us to calculate nonlinear corrections to classical linear results. 
To demonstrate our theory, we revisit the prototypical problem of a finite crack in an elastic domain subject to remote stresses and show that one can solve the complete problem to an arbitrary level of accuracy. Furthermore, we explicitly show that our solution, which is uniquely determined by remote boundary conditions and free stresses on the crack lips, does not satisfy the boundary conditions proposed in \cite{bouchbinder2008weakly}.
Furthermore, we also discover that contrary to the linear case, the asymptotic solution is not universal; different remote loads may result in different stress singularities at the crack tip. We conclude by discussing the nonlinear crack solution and its relevance in certain recognized fracture mechanisms.

\section{Linear elasticity and the Kolosov-Muskhelishvili formalism}
\subsection{Linear elasticity and Airy potential}
In elasticity, a deformation is quantified by the displacement field $\mathbf{d}(\mathbf{x})$, which maps a point in an undeformed state to its new position. Importantly, note that in our formulation $\mathbf{x}$ labels material elements in a {\it{Lagrangian}} coordinate system, which coincides with the position in the undeformed state. Upon deforming the system, an element originally located at $\mathbf{x}$ is shifted to 
 $    \mathbf{x'}=\mathbf{x} + \mathbf{d}\left(\mathbf{x}\right)$.

The change in the distance between neighboring material elements is  $d\ell'^2-d\ell^2=2u_{\mu\nu} dx^\mu dx^\nu$, 
where $d\ell$ and $d\ell'$ are the distances between two infinitesimally close points before and after the deformation, and $u_{\mu\nu}$ is the strain tensor given by
    \begin{equation}
       u_{\mu\nu}=
       \frac{1}{2}\left(\partial_\mu d_{\nu}+\partial_\nu d_{\mu} + \partial_\mu d_{\lambda}\, \partial_\nu d^{\lambda}\right).
       \label{eq:NLstrain_linear}
     \end{equation}
In what follows we are considering homogeneous and isotropic Hookean elastic solids. This framework assumes small deformations, meaning that stresses are linearly proportional to strains 
\begin{equation}
    \sigma^{\mu\nu} = {\A}^{\mu\nu\rho\sigma} u_{\rho\sigma}\;,
    \label{eq:constrel}
\end{equation}
and the elastic tensor ${\A}$, which encodes material properties and symmetries, depends only on two parameters, e.g. Poisson's ratio $\nu$ and Young's modulus $Y$. 

Furthermore, linear elasticity theory assumes small displacement gradients and thus omits terms quadratic in $\partial d$
\begin{equation}
    u_{\mu\nu}^{\left(1\right)}\approx \frac{1}{2}\left(\partial_\mu d_{\nu}+\partial_\nu d_{\mu} \right).
    \label{eq:strain_linear}
\end{equation}
The superscript $^{(i)}$ stands for $i$'th order. In equilibrium, the net force on each material element vanishes, manifested by $\boldsymbol{\sigma}$ being divergence-free as imposed by \eqref{eq:Equilibrium}, which in coordinates, and assuming \eqref{eq:strain_linear}, takes the form $\partial_\mu \sigma^{\mu\nu} = 0$.
According to Airy's formalism, the solution to \eqref{eq:Equilibrium} is represented in terms of a single stress function  $\chi^{\left(1\right)}$ \cite{airy1862strains,gurtin1963generalization},
\begin{equation}
    \sigma^{\alpha\beta \left(1\right)}= \varepsilon^{\alpha\mu} \varepsilon^{\beta\nu}\partial_{\mu\nu} \chi^{\left(1\right)}\;,
    \label{eq:Airy_potential}
\end{equation}
where $\varepsilon$ is the anti-symmetric Levi-Civita symbol \cite{thorne1973gravitation}. The stress function $\chi^{(1)}$ is determined by enforcing a geometric compatibility condition reflecting the differential relation between the strain and displacement fields 
\begin{equation}
    \varepsilon^{\alpha\mu} \varepsilon^{\beta\nu}\partial_{\alpha\beta}u_{\mu\nu}^{\left(1\right)} = 0\;.
    \label{eq:stress}
\end{equation}
Upon extracting the strain from the constitutive relation \eqref{eq:constrel} and expressing the stress using \eqref{eq:Airy_potential}, the compatibility condition reduces to the biharmonic equation for the stress function as in \eqref{eq:Rep}.

\subsection{Kolosov-Muskhelishvili formalism}
In complex coordinates, where $z = x + i y $, the Laplace operator takes the form $\nabla^2 = \partial_{z\bar{z}}$. Therefore, real harmonic functions can be represented as the real part of an analytic function, and similarly, biharmonic functions, which are solutions of \eqref{eq:Rep}, are represented as \cite{muskhelishvili_problems},
\begin{equation}
    \chi^{\left(1\right)}=\Re\{\Bar{z}\phi^{\left(1\right)}\left(z\right)+\eta^{\left(1\right)}\left(z\right)\}= \frac{1}{2}\left[\Bar{z}\phi^{\left(1\right)}\left(z\right)+z\overline{\phi^{\left(1\right)}\left(z\right)}+\eta^{\left(1\right)}\left(z\right)+\overline{\eta^{\left(1\right)}\left(z\right)}\right]
    \label{eq:Stress_Function1}
\end{equation}
where both $\phi^{\left(1\right)}\left(z\right)$ and $\eta^{\left(1\right)}\left(z\right)$ are analytic functions, and the overbar denotes the complex conjugate operator. We see that in this formalism, often referred to as the \KM formalism, the force balance equation, as well as the biharmonic equation, are identically satisfied. Therefore, it remains to determine $\phi^{\left(1\right)}(z)$ and $\eta^{\left(1\right)}(z)$ by requiring appropriate boundary conditions. 
Upon defining $\psi^{\left(1\right)}\left(z\right)=\eta'^{\left(1\right)}\left(z\right)$, \KM formulas for the stress components are 
\cite{muskhelishvili_problems,kolosov1909application}:
\begin{subequations}\label{eq:KM}
\begin{gather}
    \sigma^{xx\left(1\right)}+\sigma^{yy\left(1\right)}=2\left[\phi'^{\left(1\right)}\left(z\right)+\overline{\phi'^{\left(1\right)}\left(z\right)}\right]=4\Re\{\phi'^{\left(1\right)}\left(z\right)\}\\
    \sigma^{yy\left(1\right)}-\sigma^{xx\left(1\right)}+2i\sigma^{xy\left(1\right)}=2\left[\Bar{z}\phi''^{\left(1\right)}\left(z\right)+\psi'^{\left(1\right)}\left(z\right)\right]
\end{gather}
\end{subequations}

To complete the solution in a domain $\Omega$ one should enforce boundary conditions on $\partial\Omega$, which in a stress controlled setup reads
\begin{equation}
    \boldsymbol{\sigma}\cdot \hat{n} \rvert_{\partial\Omega}=\Vec{\tau}
    \label{eq:linearBC}
\end{equation}
with $\hat{n}$ being a normal unit vector and $\Vec{\tau}$ represents traction forces on $\partial\Omega$. 
\KM equations form the basis for solving multiple problem in elasticity, charachterized by complicated geometry, singular behaviour (such as cracks) and multiply connected domains by implementing methods from complex analysis \cite{afek2005void}. This is similar to the use of complex analysis, and in particular conformal maps, in solving Laplace equation in the complex domain.

\section{Extension to nonlinear elasticity}
A nonlinear extension of Airy's stress function approach, given with details in this section, is transparently derived within a geometric formulation of elasticity developed in \cite{ciarlet2005introduction, koiter1966nonlinear, efrati2009elastic, kupferman2015metric}. 
In this formalism, the strain tensor is defined by $u=\frac{1}{2}\left(\g-\gbar\right)$ where the reference and actual metrics $\gbar$ and $\g$ represent the rest and actual distances between material elements. If the reference metric is Euclidean, that is a stress-free configuration exists, then the definition of strain reduces to its classical nonlinear definition \eqref{eq:NLstrain_linear}. The nonlinear terms of the strain are extremely important when the displacement gradients are larger than one, even if the strain itself is small. It occurs in systems that experience large rotations, such as cracks \cite{bouchbinder2008weakly} and meta-materials\cite{Bar-Sinai2020}. Within this approach, the force balance equation that generalizes \eqref{eq:Equilibrium} is nonlinear and implicit, with the differential operator being dependent on the yet unknown actual configuration \cite{efrati2009elastic}

\begin{equation}
    {\nabla}_\mu\sigma^{\mu \nu}+\left({\Gammabar}_{\lambda \mu}^{\lambda}-{\Gamma}_{\lambda \mu}^{\lambda}\right)\sigma^{\mu\nu }=0\;,
    \label{eq:NonlinearDiv}
\end{equation}

where $\Gamma$ and ${\Gammabar}$ are the Christoffel symbols associated with $\g$ and $\gbar$, and ${\nabla}$ is the covariant derivative with respect to $\g$, that is \cite{docarmo2016differential}

\begin{equation}
    {\nabla}_\mu\sigma^{\mu \nu}=\partial_\mu \sigma^{\mu \nu}+{\Gamma}_{\mu \beta}^{\mu}\sigma^{\beta \nu}+{\Gamma}_{\mu \beta}^{\nu}\sigma^{\beta \mu}\;.
\end{equation}
Surprisingly, a representation of the nonlinear stress in terms of a scalar function still exists, namely \cite{moshe2014plane,moshe2015elastic,Bar-Sinai2020},
\begin{equation}
    \sigma^{\mu \nu}=\frac{1}{\sqrt{\det{\g} \, \det{\gbar}}}\varepsilon^{\mu \rho}\varepsilon^{\nu \sigma}\nabla_{\rho\sigma} \chi.
    \label{eq:nonlinear_stress}
\end{equation}
\eqref{eq:nonlinear_stress} is implicit since it is expanded in terms of the unknown metric $\g$, reflecting the implicit form of \eqref{eq:NonlinearDiv}. Similar to Airy's stress function, the nonlinear stress function $\chi$ is determined by imposing a geometric compatibility condition. 
Since the actual metric $\g$ describes distances in the Euclidean plane, which is flat, the compatibility condition is the vanishing of the Gaussian curvature
\begin{equation}
K(\g)=0    
\end{equation}
where the Gaussian curvature is obtained from the metric \cite{docarmo2016differential} by
\begin{equation}
 K(\g)=\frac{1}{2}g^{\alpha\beta}g^{\gamma\delta}R_{\alpha\beta\gamma\delta}   
\end{equation}
and $R_{\alpha\beta\gamma\delta}$ the Riemann curvature tensor
\begin{equation}
    R_{\alpha\beta\gamma\delta}=\frac{1}{2}\left(\partial_{\beta\gamma}g_{\alpha\delta}+\partial_{\alpha\delta}g_{\beta\gamma}-\partial_{\alpha\gamma}g_{\beta\delta}-\partial_{\beta\delta}g_{\alpha\gamma}\right)
    +g_{\mu\nu}\left(\Gamma^{\mu}_{\beta\gamma}\Gamma^{\nu}_{\alpha\delta}-\Gamma^{\mu}_{\beta\delta}\Gamma^{\nu}_{\alpha\gamma}\right)\;.
\end{equation}
In the case of small deformations relative to a Euclidean reference metric $\gbar$, this condition coincides with \eqref{eq:stress}. 
The implicit form of \eqref{eq:nonlinear_stress} prevents a direct implementation of the condition $K(\g)=0$, hence an exact analog of the biharmonic equation for the fully nonlinear problem is still lacking. To avoid this difficulty, a perturbative approach is invoked under the assumption that the nonlinear stress function can be expanded in powers of a formal small parameter $\delta$: $\chi = \chi^{(1)} + \chi^{(2)} + \dots$ with the small parameter encoded in the $n^\mathrm{th}$ order stress function $\chi^{(n)} \propto \delta^n$. In the same way, the stress can be written as
\begin{equation}
    \sigma= \sigma^{\left(1\right)} + \sigma^{\left(2\right)}+ \sigma^{\left(3\right)}+\dots
\end{equation}
In order to determine the importance of n'th order corrections of the stress, one must examine the magnitude of $\sigma^{\left(n\right)}$.

Generally, one can determine the small parameter retrospectively, but in the case of an imposed remote stress $\sigma$, the small parameter is $\delta=\sigma/Y$ where $Y$ is Young's modulus.
If the reference metric $\gbar$ is flat (in any desired coordinates system), enforcing the geometric compatibility condition, namely $K\left(g\right)=0$, order by order yields
\begin{equation}
    \begin{split}
        \Deltabar\Deltabar\chi^{(1)}&=0\;,\\
        \Deltabar\Deltabar\chi^{(2)}&=F_2\left(\chi^{(1)}\right)\;,\\
        \Deltabar\Deltabar\chi^{(3)}&=F_3\left(\chi^{(1)}, \chi^{(2)}\right)\;,\\
        &\vdots
    \end{split}
    \label{eq:nonlinearISF}
\end{equation}

Here $\Deltabar = \tfrac{1}{\sqrt{\gbar}} \partial_\mu \left(\sqrt{\gbar} \gbar^{\mu\nu} \partial_\nu\right)$ is the Laplace operator with respect to the reference metric $\gbar$, and $F_n$ are known (nonlinear) functions of stress functions of orders lower than $n$. All the equations are explicit from this point, thanks to the perturbative approach, enabling us to advance analytically.

\eqref{eq:nonlinearISF} provide a prescription for solving the nonlinear elastic problem to an arbitrary order of accuracy. The solution for $\chi^{(n)}$ consists of a (homogeneous) generic biharmonic function and a particular solution solving the non-homogeneous equation. Again, the homogeneous parts are determined by boundary conditions, as in the linear case.
The main difficulty arises from the nonlinearity of $F_n$, which prevents analytical progress. The following section shows how the complex formulation of this nonlinear formalism enables significant analytical progress.

\section{nonlinear extension of \KM formalism}
In this section, we Generalize the \KM complex formulation of Airy's stress-function approach. For that purpose, we reformulate \eqref{eq:nonlinearISF} together with its accompanying boundary conditions in complex coordinates. 
As before, the solutions of the first order stress function is the most general real biharmonic function given by \eqref{eq:Stress_Function1}.
 
For $n>1$, the solution of the $n$'th order stress-function consists of the sum of a biharmonic solution and a particular solution to \eqref{eq:nonlinearISF} denoted by $\widetilde{F}_n$
\begin{equation}
\begin{split}
    \chi^{\left(n\right)}\left(z,\bar{z}\right)&=\Re\{\Bar{z}\phi^{\left(n\right)}\left(z\right)+\eta^{\left(n\right)}\left(z\right)\}\\&+\widetilde{F}_n
 \left(\chi^{\left(1\right)},...,\chi^{\left(n-1\right)}\right)
\end{split}\label{eq:chi_n}
\end{equation} 
The analytic functions $\phi^{(n)}, \eta^{(n)}$ can be determined once boundary conditions are specified, hence expressions for the stresses in terms of $\chi^{(n)}$ are required. 

We derive the expressions for the stresses and generalize the \KM equations (\ref{eq:KM}) to the nonlinear setting 
\begin{subequations}
    \begin{align}
    &\sigma^{xx}+\sigma^{yy}=\sum_{n=1} \left[4\Re\{\phi'^{\left(n\right)}\left(z\right)\}+\Sigma_{(n)} \left(\chi^{\left(1\right)}, \dots,\chi^{\left(n-1\right)} \right)\right]\;,\\
    &\tfrac{1}{2}(\sigma^{yy}-\sigma^{xx})+i\sigma^{xy}=\sum_{n=1}\left[\Bar{z}\phi''^{\left(n\right)}\left(z\right)+\psi'^{\left(n\right)}\left(z\right)+\widetilde{\Sigma}_{(n)}  \left(\chi^{\left(1\right)},\dots,\chi^{\left(n-1\right)} \right)\right]\;.
\end{align}
\label{eq:KM2}
\end{subequations}

where ${\Sigma}_{(n)},\widetilde{\Sigma}_{(n)}$ are known functions that represent the contribution of stress functions of order smaller than $n$ to the stress field at order $n$. Therefore, ${\Sigma}_{(1)},\;\widetilde{\Sigma}_{(1)}=0$.
Although our method can be applied to any order in the expansion, to demonstrate this rather abstract formalism we focus on the second order solution
\begin{equation}
    \begin{split}
        \chitwo\left(z,\bar{z}\right)=\Re\{\Bar{z}\phitwo\left(z\right)+\etatwo\left(z\right)\}+\widetilde{F}_2\left(\chione\right)\;.
    \end{split}
 \label{second_stress_function}
\end{equation}
The particular solution for the nonlinear and non-homogeneous correction of \eqref{eq:nonlinearISF}
is analytically solvable in closed form and reads

\begin{equation}
\begin{split}
    \widetilde{F}_2\left(\chione\right) &= c_1 \left| \phione\left(z\right)\right|^2+ c_2 \left| \psione\left(z\right)  + \overline{z} \phione{}'\left(z\right)\right|^2\;,
\end{split}
\label{eq:InHomoSol}
\end{equation}

where $c_{1,2}$ are numerical factors that depend on material properties and are given in \appref{app:secondISF}. The particular solution $\tilde{F}_2$ contributes to the RHS of \eqref{eq:KM2} via the RHS of \eqref{eq:chi_n}, and thus induces effective traction forces that determine the second order functions $\phi^{(2)}\left(z\right),\eta^{(2)}\left(z\right)$. 

\section{Examples}
In this section we solve two prototypical problems in 2D elasticity using the nonlinear approach, namely a circular hole and a finite crack subjected to constant remote stresses. 
To emphasize the generality of the formalism 

in each example we implement the nonlinear generalization of \KM equations (\ref{eq:KM2}) using a different technique. 
While the circular hole problem is presented mostly for illustrating the method, the crack case is of physical importance as it questions the validity of previously reported analysis of a similar problem \cite{bouchbinder2008weakly}. 

\subsection{Circular Hole Under Stress}

\begin{figure}
     \centering
     \includegraphics[width=0.7\textwidth]{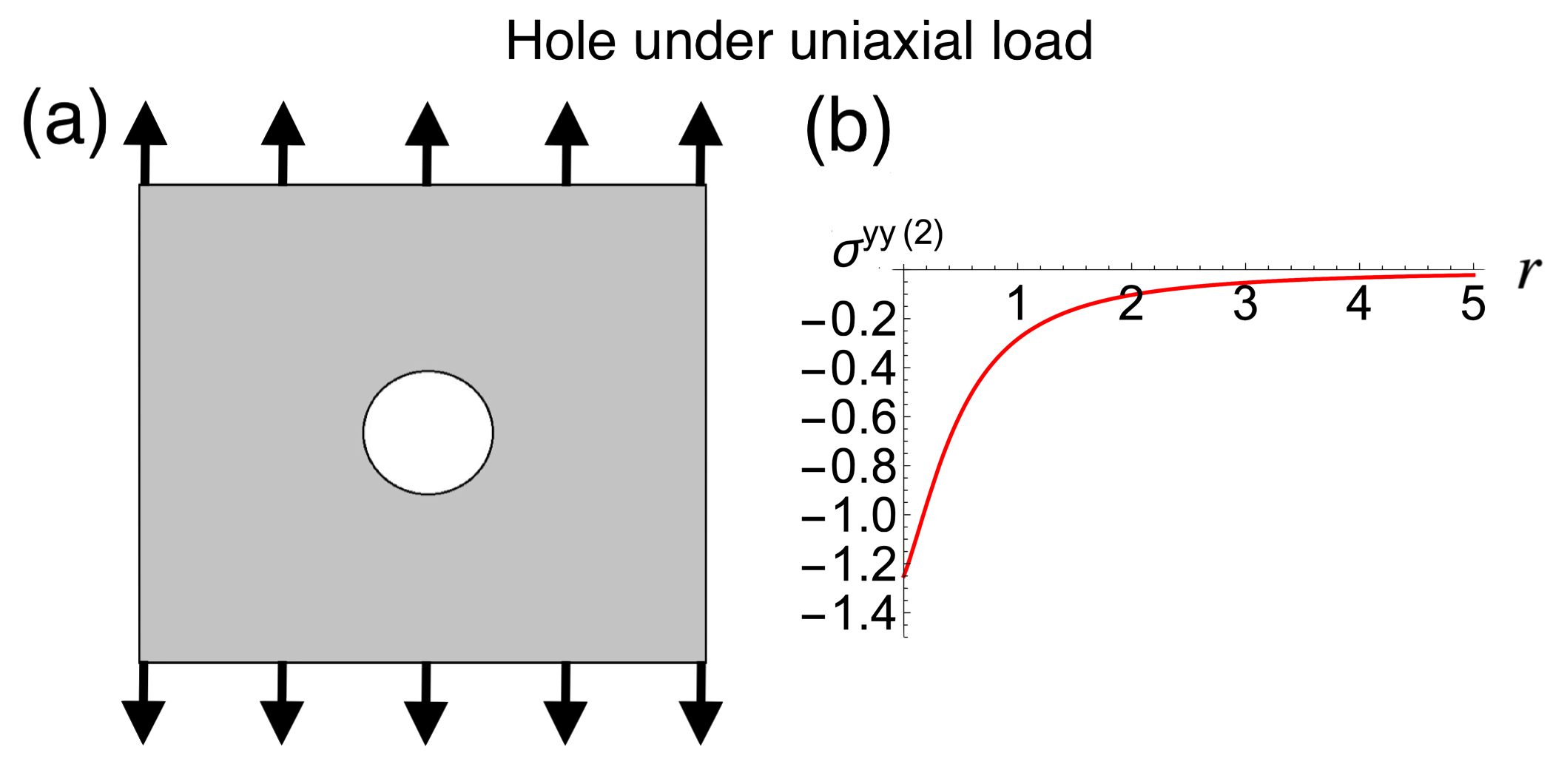}
    \caption{
    (a) Illustration of a circular hole in a 2D elastic material subjected to remote uniaxial stress.
    (b) $\sigma^{yy\left(2\right)}$ for a circular hole under uniaxial load along the x axis, with zero Poisson's ratio and Y=1. $r=1$ is the boundary of the hole.
    }
    \label{fig:hole}
\end{figure}
Consider a circular hole of radius $R=1$ in a 2D elastic material subjected to remote uniaxial stress $\sigma^{yy}_{\infty}=\sigma_\infty$ $\Bigl($see \figref{fig:hole})(a)$\Bigr)$; a problem that was first solved by Kirsch \cite{kirsch1898theorie}, and later by Kolosov \cite{kolosov1909application} and  Muskhelishvili \cite{muskhelishvili_problems} within the linear approximation.
We express the normal traction force on the hole boundary as $\tau = \tau_x+i\tau_y$. 
Upon expressing $\tau$ in terms of $\phi$ and $\eta$ via Eqs. (\ref{eq:KM})-(\ref{eq:linearBC}) the linearized boundary condition reads \cite{broberg1999cracks}:   
\begin{equation}
    \tau_x^{(1)}+i \tau_{y}^{(1)}=-i\frac{d}{d\theta}\left[\phione{}\left(z\right)+z\overline{\phione{}'\left(z\right)}+\overline{\psione{}\left(z\right)}\right]
    \label{eq:first_BC}
\end{equation}
where $z=\exp\left(i\theta\right)$ and $\theta$ the angle along the circular boundary. In the case of traction free boundary conditions on the hole, \eqref{eq:first_BC} reduces to
\begin{equation}\label{eq:traction.free}
    \phione\left(z\right)+z\overline{\phione{}'\left(z\right)}+\overline{\psione{}\left(z\right)}=const\;.
\end{equation}
The solution that satisfies both zero normal stress on the hole boundary \eqref{eq:traction.free}, and the uniform stresses at infinity, is given by
\begin{equation}
\begin{split}
    \phi^{(1)}\left(z\right)&=\frac{\sigma_\infty}{2}\left(\frac{z}{2}-\frac{1}{z}\right)\\
    \psi^{(1)}\left(z\right)&=\frac{\sigma_\infty}{2}  \left(z-\frac{1}{z}-\frac{1}{z^3}\right)
\end{split}
\label{eq:LinearHole}
\end{equation}
The stresses derived from this solution according to Eqs. (\ref{eq:KM}) are given in \appref{app:circularLinear}.

To solve higher order stress functions it is required to extend the boundary conditions \eqref{eq:first_BC} to the nonlinear setup. Indeed, in analogy to deriving \eqref{eq:first_BC} from the linear expression for the stress field \eqref{eq:KM}, we express the nonlinear stress field in terms of complex variables, using \eqref{eq:KM2}
\begin{equation}
    \tau_x+i\tau_y = -i\frac{d}{d\theta}\sum_{n}\left[\phi^{\left(n\right)}\left(z\right)+z\overline{\phi'^{\left(n\right)}\left(z\right)}+\overline{\psi^{\left(n\right)}\left(z\right)}\right] +\sum_{n}\tilde{\tau}^{(n)}\left(\chi^{\left(1\right)},\dots,\chi^{\left(n-1\right)} \right)
\end{equation}
Since the first order stress function satisfies the boundary conditions, higher order terms obey zero traction. Yet, we note that the stress functions of order $<n$ contribute to an effective traction on the boundaries, to be balanced by solution at order $n\geq 2$, that is
\begin{equation}
    i\frac{d}{d\theta}\left[\phi^{\left(n\right)}\left(z\right)+z\overline{\phi'^{\left(n\right)}\left(z\right)}+\overline{\psi^{\left(n\right)}\left(z\right)}\right] =\tilde{\tau}^{(n)}\left(\chi^{\left(1\right)},\dots,\chi^{\left(n-1\right)} \right)\;.
\label{eq:NonlinearBC_Circle}
\end{equation}

The general form of $\tilde{\tau}^{\left(2\right)}$ in the case of a circular boundary (with arbitrary boundary conditions) is given in \cite{Math}.
We apply this formalism to the second order solution for the stress function, explicitly given by \eqref{second_stress_function} and \eqref{eq:InHomoSol}. In the following, we will assume for simplicity that $\nu=0$, while a solution for general Poisson's ratio appears in \appref{app:circularSecond}. We use the fact that on the hole boundary $z=e^{i \theta}$ and therefore $\bar{z}=z^{-1}$, the $2^{\rm nd}$ order boundary condition reads
\begin{equation}
    i\frac{d}{d\theta}\left[\phi^{\left(2\right)}\left(z\right)+z\overline{\phi'^{\left(2\right)}\left(z\right)}+\overline{\psi^{\left(2\right)}\left(z\right)}\right]
    =\frac{\sigma_\infty ^2 \left(-5 z^8+42 z^6-22 z^4+14 z^2+3\right)}{16 z^4 Y}\;,
\end{equation}

with the solution
\begin{equation}
\begin{split}
     \phi^{\left(2\right)}\left(z\right)&=\frac{\sigma_\infty^2}{16 Y}\left[ z+\frac{5 }{z}+\frac{1}{z^3}\right]\\
      \psi^{\left(2\right)}\left(z\right)&=\frac{9 \,\sigma_\infty^2}{16 Y}\left[ z+\frac{20}{9 z}-\frac{1}{z^3}+\frac{4}{9 z^5}\right]
      \label{eq:NLhole}
\end{split}
\end{equation}

This form of the solution recover our formal small parameter to be $\delta = \sigma_\infty / Y$ upon comparison with the first order expression. This is in accord with our small-strains assumption. Moreover, when compared with the linear solution $\phi^{(1)}$ and $\eta^{(1)}$, the nonlinear corrections $\phi^{(2)}$ and $\eta^{(2)}$ contain a higher order term in $1/z$. Consequently, the nonlinear stress field contains higher order angular modes, and higher powers of $1/r$. Specifically, the ratio between the largest value of the stress on the hole boundary and the remote stress, aka the Stress Concentration Factor $K_t = \sigma_\mathrm{max}/\sigma_\infty$, is corrected at the second order, as obtained by estimating $\sigma^{yy}$ at $r = 1$ and $\theta = 0$
\begin{equation}
    K_t = 3\left( 1 -\frac{5}{4}\frac{\sigma_\infty}{Y} \right) \;.
    \label{eq:NLSCF}
\end{equation}
\figref{fig:hole}(b) shows the ratio between the second order correction of $\sigma^{yy}$ and the first order solution i.e., $\delta\sigma^{yy\left(2\right)}$  along the $\theta=0$ line.
We emphasize that the solutions \eqref{eq:NLhole} and \eqref{eq:NLSCF} are valid for $\nu = 0$, and expressions for arbitrary Poisson's ratio are given in \appref{app:circularSecond}. 
\begin{figure}
     \centering
     \includegraphics[width=0.7\textwidth]{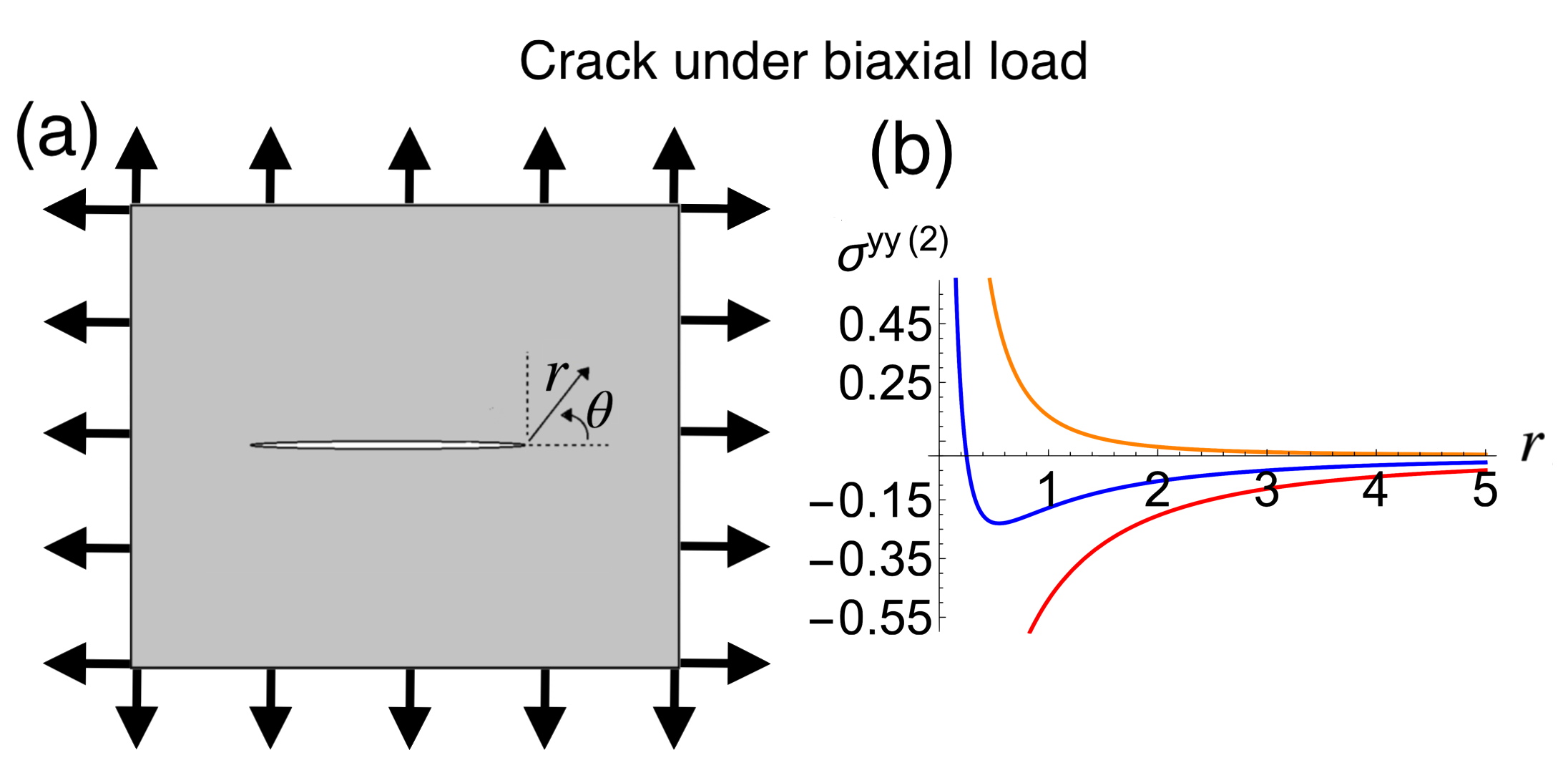}
    \caption{
    (a) Illustration of a mode I crack.
    (b) $\sigma^{yy\left(2\right)}$ of a mode I crack for isotropic stress $\sigma^{xx}_\infty=\sigma^{yy}_\infty$ (red), traceless symmetric stress $\sigma^{xx}_\infty=-\sigma^{yy}_\infty$ (orange) and a superposition of the two, $\sigma^{xx}_\infty=0$ along the crack with zero Poisson's ratio in terms of the small parameter and $Y=1$. $r=0$ is the tip of the crack.
    }
    \label{fig:crack}
\end{figure}

\subsection{Finite Straight Cracks}
In this section we consider a finite straight crack embedded in an infinite medium, and subjected to uniform remote stresses. Specifically, we solve a nonlinear mode I problem, and give the nonlinear mode II solution in \appref{app:mode2}.  
The crack problem is characterized by its geometry $C = \{(x,y): -1<x<1 \,\wedge \,y=0 \}$, by the traction forces acting at infinity $\sigma_{\infty}$, and by the traction free boundary conditions on the crack faces.
We note that the displacements, may be discontinuous across the crack. Correspondingly, it is necessary to enforce boundary conditions on the top and bottom crack faces separately.
At the linear level we follow the Westergaard method \cite{broberg1999cracks,muskhelishvili_problems}. Using \eqref{eq:KM} and defining $\Psi=\psi'$, $\Phi=\phi'$ and $\Omega=\Phi+\Psi+z\Phi'$, we find
\begin{equation}
    \sigma^{yy\left(1\right)} + i \sigma^{xy\left(1\right)}=\Omega^{\left(1\right)}\left(z\right)+\overline{\Phi^{\left(1\right)}\left(z\right)}+\left(\Bar{z}-z\right)\Phi'^{\left(1\right)}\left(z\right)
    \label{eq:CrackBCI}
\end{equation}
Upon approaching the crack from above or below, we denote $z=x \pm i\epsilon$, the normal stress in complex form reads
\begin{equation}
   \left. \left(\sigma^{yy\left(1\right)}+i\sigma^{xy\left(1\right)}\right)\right|_C=\Omega^{\left(1\right)}\left(x\pm i \epsilon\right)+\overline{\Phi^{\left(1\right)}}\left(x\mp i \epsilon\right)\;,
    \label{eq:ComplexCrackBCI}
\end{equation}
where the third term in \eqref{eq:CrackBCI} vanishes as $\epsilon \to 0$.

Adding and subtracting the upper and lower limits of \eqref{eq:ComplexCrackBCI}, and defining $\Omega^{\left(1\right)}\left(z\right) = \tfrac{1}{2}\left[I^{\left(1\right)}\left(z\right) + J^{\left(1\right)}\left(z\right)\right]$ and $\overline{\Phi^{\left(1\right)}}\left(z\right) = \tfrac{1}{2}\left[I^{\left(1\right)}\left(z\right) - J^{\left(1\right)}\left(z\right)\right]$, yields
\begin{equation}
\begin{split}
   0 = I^{\left(1\right)}\left(x+ i \epsilon\right)+I^{\left(1\right)}\left(x- i \epsilon\right)\\
   0 = J^{\left(1\right)}\left(x+ i \epsilon\right)-J^{\left(1\right)}\left(x- i \epsilon\right)\;,
\end{split}
\label{eq:westergaard.faces}
\end{equation}
where the left hand side vanished due to traction free boundary conditions. These equations for $I^{\left(1\right)}\left(z\right)$ and $J^{\left(1\right)}\left(z\right)$ replace the equations for $\Omega^{\left(1\right)}\left(z\right), \Phi^{\left(1\right)}\left(z\right)$, and correspondingly for $\psi^{\left(1\right)}\left(z\right),\phi^{\left(1\right)}\left(z\right)$. Once the equations for $I^{\left(1\right)}\left(z\right)$ and $J^{\left(1\right)}\left(z\right)$ are solved, we can directly derive the solution for the stress function and the stress fields.

Soon we will see that the form of the equation for $J^{\left(1\right)}\left(z\right)$ in \eqref{eq:westergaard.faces} is analytically solvable. Therefore, we would prefer to rewrite the equation for $I^{\left(1\right)}\left(z\right)$ in the same form as that for $J^{\left(1\right)}\left(z\right)$.
For that, we define $L^{\left(1\right)}\left(z\right)=G\left(z\right)I^{\left(1\right)}\left(z\right)$ with $G\left(z\right)$ an analytic function that satisfies on the crack boundary $G( x + i \epsilon) = -G( x - i \epsilon)$.
Indeed, we note that the specific choice $G\left(z\right)=\sqrt{\left(z+1\right)\left(z-1\right)}$ satisfies this condition, and a direct substitution in \eqref{eq:westergaard.faces} yields
\begin{equation}
\begin{split}
   L^{\left(1\right)}\left(x+ i \epsilon\right)-L^{\left(1\right)}\left(x- i \epsilon\right)&=0\\
   J^{\left(1\right)}\left(x+ i \epsilon\right)-J^{\left(1\right)}\left(x- i \epsilon\right)&=0
\end{split}
\label{eq:Fjump}
\end{equation}
Since $L^{\left(1\right)}\left(z\right)$ and $J^{\left(1\right)}\left(z\right)$ have the same value on both sides of the branch cut, they are analytic everywhere.

For the non-homogeneous version of \eqref{eq:Fjump} one can use the Sokhotski-Plemelj theorem. 

From the relation $L^{\left(1\right)}(z) = G(z) I^{\left(1\right)}(z)$ and from the relation between $I^{\left(1\right)}\left(z\right),\;J^{\left(1\right)}\left(z\right)$ and $\Phi^{\left(1\right)}\left(z\right), \Omega^{\left(1\right)}\left(z\right)$ we find 
\begin{equation}
\begin{split}
    \Omega^{\left(1\right)}\left(z\right)&=\frac{1}{2}\left[\frac{L^{\left(1\right)}\left(z\right)}{\sqrt{z^2-1}}+J^{\left(1\right)}\left(z\right)\right]\\
    \overline{\Phi^{\left(1\right)}}\left(z\right)&=\frac{1}{2}\left[\frac{L^{\left(1\right)}\left(z\right)}{\sqrt{z^2-1}}-J^{\left(1\right)}\left(z\right)\right]
\end{split}
\label{eq:sol_crack_1st}
\end{equation}
{Since $L^{\left(1\right)}\left(z\right)$ and $J^{\left(1\right)}\left(z\right)$ are analytic functions, the celebrated $1/\sqrt{r}$ singularity at the crack tip can already be seen by direct substitution in \eqref{eq:CrackBCI}}.
The analytic functions $L^{\left(1\right)}\left(z\right)$ and $J^{\left(1\right)}\left(z\right)$ are determined by enforcing the boundary conditions on the remote stresses at infinity, as well as the continuity of the elastic fields at $y=0$ out of the crack domain. 
In the case of Mode I crack, where the stress at infinity is 
\begin{equation}
   \sigma_\infty=\left( 
    \begin{array}{cc}
        \sigma^{xx}_{\infty} & 0 \\
        0 & \sigma^{yy}_\infty \\
    \end{array}
\right)\;,
\end{equation}
the solution for $L^{\left(1\right)}\left(z\right)$ and $J^{\left(1\right)}\left(z\right)$ is given by

\begin{equation}
\begin{split}
    L^{\left(1\right)}\left(z\right)&=\sigma^{yy}_\infty \;z\\
        J^{\left(1\right)}\left(z\right)&=\frac{\sigma^{yy}_\infty-\sigma^{xx}_\infty}{2}\;.
        \end{split}
        \label{eq:LinearAnalyticPsol}
\end{equation}

To solve the second order stress function we first generalize \eqref{eq:CrackBCI} to the nonlinear setup. 
For that we derive a nonlinear generalization of \eqref{eq:CrackBCI}  using the nonlinear \KM equations (\ref{eq:KM2})
\begin{equation}
\begin{split}
    \sigma^{yy} + i \sigma^{xy}=&\sigma^{yy\left(1\right)} + i \sigma^{xy\left(1\right)}
    +\sum_{n=2}\left(\Omega^{\left(n\right)}\left(z\right)+\overline{\Phi^{\left(n\right)}\left(z\right)}+\left(\Bar{z}-z\right)\Phi'^{\left(n\right)}\left(z\right)\right)\\
    &+\sum_{n=2}\left[\tfrac{1}{2}\Sigma_{(n)}  \left(\chi^{\left(1\right)},\dots,\chi^{\left(n-1\right)} \right)+\tilde{\Sigma}_{(n)}  \left(\chi^{\left(1\right)},\dots,\chi^{\left(n-1\right)}  \right)\right]\;.
    \end{split}
\end{equation}
We denote
\begin{equation}
    \begin{split}
        \Delta \sigma^{yy\left(n\right)}&=\Re{\tfrac{1}{2}\Sigma_{(n)}  \left(\chi^{\left(1\right)},\dots,\chi^{\left(n-1\right)} \right)+\tilde{\Sigma}_{(n)}  \left(\chi^{\left(1\right)},\dots,\chi^{\left(n-1\right)} \right)}\\
        \Delta \sigma^{xy\left(n\right) }&=\Im{\tfrac{1}{2}\Sigma_{(n)}  \left(\chi^{\left(1\right)},\dots,\chi^{\left(n-1\right)} \right)+\tilde{\Sigma}_{(n)}  \left(\chi^{\left(1\right)},\dots,\chi^{\left(n-1\right)} \right)}
    \end{split}
    \label{eq:DeltaSigma2}
\end{equation}
and interpret them as the $n$'th order tensile and shear stresses induced by lower order solutions.

As in the circular hole case, the first order solution for the stress function balances the imposed boundary conditions. Therefore, higher order stresses should identically vanish on the boundaries. Specifically, upon requiring zero traction forces on the crack faces we find at all orders
\begin{equation}
\begin{split}
   \Omega^{\left(n\right)}\left(x\pm i\epsilon\right)+\overline{\Phi^{\left(n\right)}}\left(x\mp i\epsilon\right) =-\Delta \sigma^{yy\left(n\right)} \left(x\pm i\epsilon\right) - i \Delta \sigma^{xy\left(n\right)}\left(x\pm i\epsilon\right)\
    \end{split}\label{eq:CrackBCII}
\end{equation}

We express the boundary conditions at second order by a direct substitution of the first order solution in \eqref{eq:DeltaSigma2} with $n=2$ on the crack faces and find 
\begin{equation}
\begin{split}
    \Delta\sigma^{yy\left(2\right)}\left(x\pm i \epsilon\right)=&\xi_1 (\nu) \frac{\left(\sigma^{xx}_{\infty}\right)^2-2\sigma^{xx}_{\infty}\sigma^{yy}_{\infty}-3\left(\sigma^{yy}_{\infty}\right)^2}{Y}\\
    i\Delta\sigma^{xy\left(2\right)}\left(x\pm i \epsilon\right)=&\pm\frac{\sigma^{yy}_{\infty} \left( \sigma^{xx}_{\infty}-\sigma^{yy}_{\infty}\right)}{Y \left(x^2-1\right)^{3/2}}\left[\xi_2(\nu) x+\xi_3(\nu) x^3\right]\;,
\end{split}
\label{eq:deltastress2}
\end{equation}
where $\xi_i$ are numerical factors that depend on Poisson's ratio (see \appref{app:mode1}).
As before, our goal is to solve for $\Omega^{\left(2\right)}\left(z\right)$ and $\Phi^{\left(2\right)}\left(z\right)$. For that we repeat the technique from the first order analysis. Upon adding and subtracting the upper and lower limits of \eqref{eq:CrackBCII}, and defining $\Omega^{\left(2\right)}\left(z\right) = \tfrac{1}{2}\left[I^{\left(2\right)}\left(z\right) + J^{\left(2\right)}\left(z\right)\right]$ and $\overline{\Phi^{\left(2\right)}}\left(z\right) = \tfrac{1}{2}\left[I^{\left(2\right)}\left(z\right) - J^{\left(2\right)}\left(z\right)\right]$, we find
\begin{equation}
\begin{split}
   I^{\left(2\right)}\left(x+ i \epsilon\right)+I^{\left(2\right)}\left(x- i \epsilon\right)=-2\Delta\sigma^{yy}{}^{\left(2\right)}\left(x+i \epsilon\right)\\
   J^{\left(2\right)}\left(x+ i \epsilon\right)-J^{\left(2\right)}\left(x- i \epsilon\right)=-i\left[\Delta\sigma^{xy}{}^{\left(2\right)}\left(x+i \epsilon\right)-\Delta\sigma^{xy}{}^{\left(2\right)}\left(x-i \epsilon\right)\right]
\end{split}
\label{eq:ComplexCrackBCII}
\end{equation}
At this stage, upon defining $L^{(2)}\left(z\right) = G(z) I^{(2)}(z)$, we can transform the first equation to the form of the second one, similar to the transition from \eqref{eq:westergaard.faces} to \eqref{eq:Fjump}. Instead, here we perform a stronger transformation with which the equations become homogeneous
\begin{equation}
\begin{split}
G_2\left(z\right)&=\pm\frac{\sigma^{yy}_{\infty} \left( \sigma^{xx}_{\infty}-\sigma^{yy}_{\infty}\right)}{Y \left(z^2-1\right)^{3/2}}\left(\xi_2 z+\xi_3 z^3\right)\\
     L^{\left(2\right)}\left(z\right)&= G\left(z\right)\left[I^{\left(2\right)}\left(z\right)+\Delta\sigma^{yy\left(2\right)}\right]\\
      M^{\left(2\right)}\left(z\right)&= G_2\left(z\right) + J^{\left(2\right)}\left(z\right)\;,
    \end{split}
    \label{eq:2ndtransform}
\end{equation}
where $G_2\left(z\right)$ is a non-homogeneous solution for $J^{(2)}\left(z\right)$ as reflected from the complex continuation of $\Delta\sigma^{xy\left(2\right)}$ in \eqref{eq:deltastress2}.
Indeed, we identify $i\Delta\sigma^{xy\left(2\right)}\left(x\pm i \epsilon\right)=G_2\left(x\pm i\epsilon\right)$, with which Eqs. (\ref{eq:ComplexCrackBCII}) now read
\begin{equation}
\begin{split}
   L^{\left(2\right)}\left(x+i\epsilon\right)- L^{\left(2\right)}\left(x-i\epsilon\right)&=0\\
    M^{\left(2\right)}\left(x+i\epsilon\right)- M^{\left(2\right)}\left(x- i \epsilon\right)&=0
\end{split}
\label{eq:ComplexCrackBCIIHom}
\end{equation}
The choice of $L^{\left(2\right)}\left(z\right)$ and $M^{\left(2\right)}\left(z\right)$ transformed \eqref{eq:ComplexCrackBCII} into a set of homogeneous equations. In a case where such a choice is not possible, one may use Sokhotski-Plemelj theorem.

It follows from \eqref{eq:ComplexCrackBCIIHom} that the second order correction to the stress function is
\begin{equation}
\begin{split}
    \Omega^{\left(2\right)}\left(z\right)=\frac{1}{2}\left[-G_2\left(z\right)+\frac{L^{\left(2\right)}\left(z\right)}{\sqrt{z^2-1}}+M^{\left(2\right)}\left(z\right)-\Delta\sigma^{yy\left(2\right)}\right]\\
    \overline{\Phi^{\left(2\right)}}\left(z\right)=\frac{1}{2}\left[G_2\left(z\right)+\frac{L^{\left(2\right)}\left(z\right)}{\sqrt{z^2-1}}-M^{\left(2\right)}\left(z\right)-\Delta\sigma^{yy\left(2\right)}\right]\;,
\end{split}
\end{equation}
where $L^{\left(2\right)}\left(z\right)$ and $M^{\left(2\right)}\left(z\right)$ are analytic functions determined by the remote boundary conditions and the displacement continuity outside the crack. 
As in the linear case, also here we can identify the stress singularity at the crack tip regardless of the exact form of $L^{\left(2\right)}\left(z\right)$ and $M^{\left(2\right)}\left(z\right)$. Since $\Omega^{\left(2\right)}\left(z\right)$ and $\overline{\Phi^{\left(2\right)}}\left(z\right)$ are linearly dependent on $G_2\left(z\right)$, upon expanding it near the crack tip $\left(z=\pm 1\right)$, its form reveals stronger singularities, which include terms like $1/r$ and $1/r^{3/2}$.

To find the explicit form of $L^{\left(2\right)}\left(z\right)$ and $M^{\left(2\right)}\left(z\right)$ we note that the first order solution contributes to the second order stresses a uniform stress at infinity, 
\begin{equation}
   \Delta\sigma_\infty^{\left(2\right)}=\left( 
    \begin{array}{cc}
        s^{xx\left(2\right)} & 0 \\
        0 & s^{yy\left(2\right)}\\
    \end{array}
\right)\;,
\end{equation}
where the constants entries are given in \appref{app:deltastress2inf}.
The solution for $L^{\left(2\right)}\left(z\right)$ and $M^{\left(2\right)}\left(z\right)$ is
\begin{equation}
\begin{split}
    L^{\left(2\right)}\left(z\right) &= p_1 z\\
   M^{\left(2\right)}\left(z\right) &= p_2\;,
\end{split}
\end{equation}
where $p_1$ and $p_2$ are constants given by
\begin{equation}
\begin{split}
   p_1&=\left(\Delta\sigma^{yy\left(2\right)}-s^{yy\left(2\right)}\right)\\
   p_2&=\frac{1}{2}\left(s^{xx\left(2\right)}-s^{yy\left(2\right)}\right)-\frac{\left(\nu+1\right)^2}{8Y}\left(\sigma^{xx}_{\infty}-\sigma^{yy}_{\infty}\right)\sigma^{yy}_{\infty}
\end{split}
\end{equation}
and $\Delta\sigma^{yy\left(2\right)}$ is a constant given by \eqref{eq:deltastress2}.

{Interestingly, while the stress singularity at the crack tip in the linear approximation is proportional to $1/\sqrt{r}$ for any externally imposed loads, the singularity at the second order depends on the specific loading. This is seen from the explicit form of $G_2\left(z\right)$ given in \eqref{eq:2ndtransform}, which vanishes for isotropic loading $\sigma^{xx}_{\infty}=\sigma^{yy}_{\infty}$.

We deduce that the $1/r$ singularity is generic at the second order, while the stronger singularity $1/r^{3/2}$ disappears when the imposed loading is bi-axial. The second and higher order singularities become important as one approaches the crack tip. Far from the crack tip, the linear solution is clearly valid.} In \figref{fig:crack}(b), one can see the correction to $\sigma^{yy}$, namely $\sigma^{yy\left(2\right)}$ in three cases: isotropic stress, traceless symmetric stress and a simple superposition of the two where $\sigma^{xx}_\infty=0$. We see that as we approach the crack tip, the effect of the traceless symmetric stress dominates because of its stronger singularity.

{An important consequence of our analysis is the limitation on methods that aim at solving for the asymptotic behaviour at the vicinity of the crack tip. In the linear approximation, the asymptotic behaviour near the tip of the crack can be found without imposing the remote stress boundary conditions, although without determining the coefficients, such as the stress intensity factor\cite{irwin1957analysis}. Here we find that a second order solution cannot be found purely asymptotically because the remote stresses determines the order of the singularity close to the crack tip. This observation questions the validity of the asymptotic analysis presented in \cite{bouchbinder2008weakly,bouchbinder2010weakly}}.

\section{Summary and Discussions}
\label{sec:summary}

\begin{figure}
     \centering
     \includegraphics[width=\textwidth]{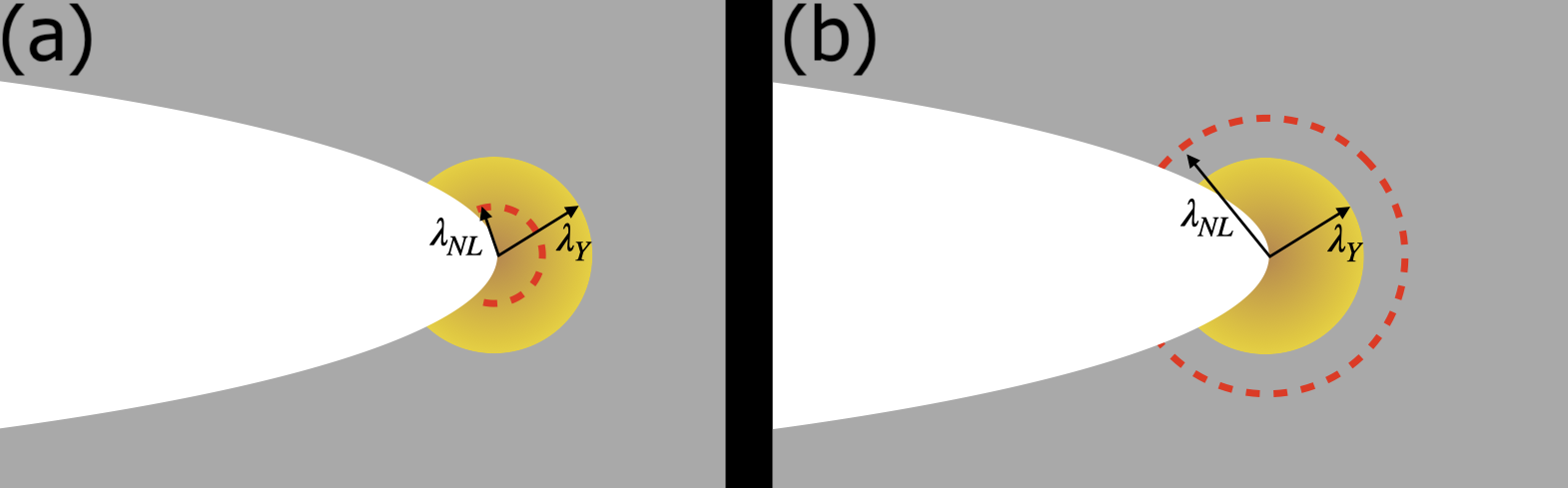}
    \caption{The importance of nonlinearity at the onset of fracture. Nonlinear contributions to the elastic fields are important in the region bounded by the red dashed line $r < \lambda_{NL}$. Plastic deformations occur inside the yellow region $r<\lambda_{Y}$. (a) When $\sigma_Y<Y$ the material fails at small strains, thus nonlinear corrections are significant only on scales smaller than the plastic zone size, and therefore irrelevant. (b) When $\sigma_Y>Y$ the material can support large strains prior to failure, hence at the onset of failure strains are sufficiently large, with non-negligible nonlinear corrections on scale larger then the process zone. }
    \label{fig:nonlinearity}
\end{figure}

In this study, based on an earlier geometric extension of the Airy stress function approach, we developed a geometrically nonlinear version of the \KM formulae for analyzing 2D elastic problems.
Our method allows us to port tools from complex analysis to study nonlinear elasticity of 2D solids, with either regular or singular boundaries.
Such problems are essential for understanding natural phenomena such as fractures in solids under large deformations. 
Motivated by fracture mechanics, we focus on the problem of a straight finite crack and use our generalized \KM equations to study the nonlinear elastic solution.
While in this work the calculations performed up to second order, our method is applicable to any desired level of accuracy.


Another exciting consequence that arises from the second-order correction to the Mode II crack, known to be unstable compared with Mode I crack, is a feature strongly related to the celebrated Principle of Local Symmetry \cite{gol1974brittle}. This phenomena coheres with our theory, where a prominent distinction between Mode I and Mode II cracks is observed: Nonlinear corrections to the Mode II stress field in the vicinity of crack tip contains contributions with symmetries characteristic of Mode I. This is contrary to nonlinear corrections to Mode I, which preserve its symmetries. This is shown in \appref{app:mode2}, and suggests a future research direction for the derivation of the principle of local symmetry within the framework of nonlinear crack analysis. In particular the relevance of Mode I contributions to seemingly pure Mode II fracture should be further studied.

It is important to note that the relevance of our work to fracture mechanics is not always obvious. 
While the nonlinear corrections to the stress field become important only very close to the crack tip, in this region, new physics such as plasticity may dominate. Therefore, judgement regarding the relevance of our results to fracture mechanics reduces to comparing two different length scales: $\lambda_Y$ the scale below which fracture occurs, and $\lambda_\text{NL}$ the scale below which nonlinear elastic corrections are non-negligible. If $\lambda_Y<\lambda_\text{NL}$ then nonlinear corrections are important. 
A dimensional analysis argument clarifies this comparison of scales. On one hand, the scale $\lambda_Y$ reflects the balance of linear stresses with the microscopic yield stress $\sigma_Y \approx \sigma^{\infty} \sqrt{a/\lambda_Y}$ where $a$ is the length of the crack, hence $\lambda_{Y} \approx a (\sigma^\infty / \sigma_Y)^2$ \cite{DUGDALE1960100}. On the other hand, the scale where nonlinear correction become non-negligible reflects equal contribution from the $1/r$ and $1/\sqrt{r}$ terms $\sigma^\infty/\sqrt{\lambda_\text{NL}/a} \approx (\sigma^\infty)^2/\left[Y \left(\lambda_\text{NL}/a\right)\right]$, hence $\lambda_\text{NL} \approx a (\sigma^\infty/Y)^2$. The conditions for nonlinear correction to be non-negligible $\lambda_Y<\lambda_\text{NL}$, reduces to $\sigma_Y>Y$. This result is reasonable, as it suggests that materials that support large strain before failure, will require nonlinear corrections. This is summerized in \figref{fig:nonlinearity}.

An important test for any new theory is the comparison with previous results. The most prominent nonlinear analysis of the asymptotic elastic fields in the vicinity of a crack was published in a series of papers \cite{bouchbinder2008weakly,bouchbinder20091,bouchbinder2010autonomy,bouchbinder2010weakly,livne2008breakdown,livne2010near}. In these works, like in our work, the authors recovered the $1/r$ and $1/r^{3/2}$ singularities but excluded the latter based on a newly suggested boundary condition at the crack tip, that is requiring the elastic force $\mathbf{f} = \mathrm{div} \sigma$, to vanish on the crack tip. This new boundary condition aimed to replace the remote boundary conditions that are absent from the asymptotic problem. 
Contrary to the asymptotic analysis, our analysis does not require any additional boundary condition other than the standard normal forces on the boundaries. Our results show that the $1/r^{3/2}$ singularity can survive, depending on the specific remote loading, and therefore show that the problem cannot be solved only asymptotically. This shows the failure of the extra boundary condition of zero force on the crack tip. To settle this disagreement, we emphasize that the bulk equation $\mathrm{div} \sigma = 0$ should hold at the interior of the solid. A non-zero value on the boundary is in no contradiction with the free boundary conditions that are reflected by $\boldsymbol{\sigma} \cdot \hat{n}=0$.
Indeed, while in the linear approximation at the crack tip, the condition $\mathrm{div} \sigma = 0$  holds, it does not hold along the crack faces, and as said, is not in contradiction with the bulk equation and boundary conditions. 
Our results, which satisfy the boundary conditions and bulk equation to first and second order, question the validity of asymptotic nonlinear crack analysis published in the past and call for revisiting the comparison with experimental results.

The nonlinear geometric approach to elasticity, which is the basis for the nonlinear \KM formalism developed in this work, has many advantages, among which is the ability to calculate nonlinear interactions between sources of stresses \cite{Bar-Sinai2020} and nonlinear residually stressed solids \cite{moshe2014plane} in the presence of large deformation-gradients. Similarly, one can use the geometrically nonlinear \KM formalism we propose to calculate the nonlinear interaction of cracks with singular sources of stresses, such as topological defects, and with the background stress in residually stressed solids, such as living growing matter. We leave this promising direction of research for future study. 

\section{Acknowledgements}
This work was supported by the Israel Science Foundation grant No. 1441/19, and by the International Research Project ``Non-Equilibrium Physics of Complex Systems'' (IRP-PhyComSys, France-Israel).

\appendix

\section{Second order ISF}\label{app:secondISF}
Our treatment of the nonlinear elastic problem is based on a perturbative solution to the nonlinear stress function, where we expand the stress function in terms of a small parameter. A flat reference metric $\gbar$ gives a set of non-homogeneous harmonic equations, in which the non-homogeneous part depends only on lower orders solutions of the stress function, as shown in \eqref{eq:nonlinearISF}. For example, the equation for the second-order correction is \cite{Math} 

\begin{multline}
    \frac{Y}{2}\Deltabar\Deltabar\chi^{\left(2\right)}\left(z,\bar{z}\right)=\left(25-22\nu+\nu^2\right)\left|\phi^{\left(1\right)''}\right|^2\\
    -\left(1+\nu\right)^2\left(2z\phi^{\left(1\right)'''}\overline{\phi^{\left(1\right)''}}+2\bar{z}\phi^{\left(1\right)''}\overline{\phi^{\left(1\right)'''}}
    +2\psi^{\left(1\right)'}\overline{\phi^{\left(1\right)'''}}+2\overline{\psi^{\left(1\right)'}}\phi^{\left(1\right)'''}+\left|\bar{z}\phi^{\left(1\right)'''}+\psi^{\left(1\right)''}\right|^2 \right)
\end{multline}
The general solution is
\begin{multline}
 \chi^{\left(2\right)}\left(z,\bar{z}\right)=\frac{1}{2}\left(\Bar{z}\phi^{\left(2\right)}\left(z\right)+z\overline{\phi^{\left(2\right)}\left(z\right)}+\eta^{\left(2\right)}\left(z\right)+\overline{\eta^{\left(2\right)}\left(z\right)}\right)\\
 -\frac{1}{8 Y}\left(-29+14\nu-5\nu^2\right)\left|\phi^{\left(1\right)}\left(z\right)\right|^2
  -\frac{1}{8 Y}\left(1+\nu\right)^2\left|\Bar{z}\phi'^{\left(1\right)}\left(z\right)+\psi^{\left(1\right)}\left(z\right)\right|^2
\end{multline}

\section{Radial Symmetry}
\subsection{Circular Hole - First Order}\label{app:circularLinear}
\eqref{eq:LinearHole} shows the linear stress function of a unit circular hole subjected to uniaxial stress $\sigma_\infty$ along the $y$ axis. The derived stresses are:
\begin{subequations}
\begin{gather}
    \sigma^{xx\left(1\right)}=\sigma_\infty\frac{1}{2r^2}\left[\cos{2\theta}+\left(2-\frac{3}{r^2}\right)\cos{4\theta}\right]\\
    \sigma^{xy\left(1\right)}=\sigma_\infty\frac{1}{2r^2}\left[-\sin{2\theta}+\left(2-\frac{3}{r^2}\right)\sin{4\theta}\right]\\
    \sigma^{yy\left(1\right)}=\sigma_\infty\left[1+\frac{3}{2r^2}\cos{2\theta}+\left(\frac{3}{2r^4}-\frac{1}{r^2}\right)\cos{4\theta}\right]
\end{gather}
\end{subequations}
where $r$ is the distance from the hole center and $\theta$ is the angle relative to the $x$ axis.

\subsection{Circular Hole - Second order correction}\label{app:circularSecond}
The homogeneous part of the second order correction to the stress function with an arbitrary Poisson's ratio is
\begin{subequations}
\begin{gather}
     \phi^{\left(2\right)}\left(z\right)=\frac{\sigma_\infty^2}{16 Y}\left[-\left(\nu-1\right)z+\frac{\left(5-2\nu+\nu^2\right)}{z}+\frac{\left(1+\nu\right)^2 }{z^3}\right]\\
      \psi^{\left(2\right)}\left(z\right)=\frac{\sigma_\infty^2}{16 Y}\left[\left(\nu-3\right)^2z+\frac{4\left(5-2\nu+\nu^2\right)}{z}-\frac{\left(\nu-3\right)^2 }{z^3}+\frac{4\left(1+\nu\right)^2 }{z^5}\right]
\end{gather}
\end{subequations}

when the full derivation appears in \cite{Math}.
From these functions one can derive the second order correction to the stress, which take the following form:
\begin{subequations}
\begin{align}
\sigma^{xx\left(2\right)}\left(r,\theta\right)&=\frac{\sigma_\infty^2}{16 Y}\left[f_1\left(r\right)+f_2\left(r\right)\cos 2 \theta+f_3\left(r\right)\cos 4 \theta+f_4\left(r\right)\cos 6 \theta \right]\\
\sigma^{xy\left(2\right)}\left(r,\theta\right)&=\frac{\sigma_\infty^2}{16 Y}\left[f_5\left(r\right)\cos 2 \theta+f_6\left(r\right)\cos 4 \theta+f_7\left(r\right)\cos 6 \theta \right]\\
\sigma^{yy\left(2\right)}\left(r,\theta\right)&=\frac{\sigma_\infty^2}{16 Y}\left[f_8\left(r\right)+f_9\left(r\right)\cos 2 \theta+f_{10}\left(r\right)\cos 4 \theta-f_4\left(r\right)\cos 6 \theta \right]
\end{align}
\label{eq:Bstress}
\end{subequations}

where the functions $f_i\left(r\right)$ are

\begin{subequations}
\begin{align}
    f_1\left(r\right)&=\frac{3 (\nu -6) \nu
   +11}{256 r^4}+\frac{(5 \nu +9) (\nu +1)}{128 r^6}-\frac{9 (\nu +1)^2}{256 r^8}\\
   f_2\left(r\right)&=\frac{(\nu -1) \nu }{32 r^2}+\frac{(14-3 \nu ) \nu -31}{128 r^4}+\frac{-9 \nu ^2+6 \nu +39}{128 r^6}+\frac{3 (\nu -3) (\nu
   +1)}{64 r^8}\\
   f_3\left(r\right)&=\frac{\nu  (5 \nu
   -2)-23}{128 r^2}+\frac{(4-3 \nu ) \nu +9}{32 r^4}+\frac{3 (\nu -3) (\nu +1)}{64 r^6}\\
   f_4\left(r\right)&=\frac{3 (\nu -3) (\nu +1)}{128 r^2}+\frac{1-3 (\nu -2) \nu
   }{32 r^4}+\frac{5 (\nu -1)^2}{64 r^6}\\
   f_5\left(r\right)&=\frac{\nu ^2-2 \nu +5}{128 r^2}-\frac{3
   \left(\nu ^2-2 \nu +13\right)}{128 r^4}-\frac{3 \left(\nu ^2-2 \nu -7\right)}{64 r^6}+\frac{3 (\nu -3) (\nu +1)}{64 r^8}\\
   f_6\left(r\right)&=\frac{\nu ^2-2 \nu -19}{128 r^2}-\frac{3 \left(\nu ^2-2 \nu -7\right)}{64 r^4}+\frac{3
   (\nu -3) (\nu +1)}{64 r^6}\\
   f_7\left(r\right)&=\frac{3 (\nu -3) (\nu
   +1)}{128 r^2}+\frac{-3 \nu ^2+6 \nu +1}{32 r^4}+\frac{5 (\nu -1)^2}{64 r^6}\\
   f_8\left(r\right)&=-\frac{9 (\nu +1)^2}{256 r^8}+\frac{3 (\nu -7) (\nu +1)}{256 r^4}+\frac{3 (\nu +5) (\nu
   +1)}{128 r^6}\\
   f_9\left(r\right)&=-\frac{3 (\nu -3) (\nu +1)}{64 r^8}+\frac{3 ((\nu -6) \nu -15)}{128 r^6}+\frac{-((\nu
   -4) \nu )-15}{64 r^2}+\frac{\nu  (3 \nu +2)+47}{128 r^4}\\
   f_{10}\left(r\right)&=\frac{-\nu -6}{16 r^4}-\frac{3 (\nu -3) (\nu +1)}{64 r^6}+\frac{\nu  (3 \nu +2)+15}{128
   r^2}
\end{align}
\end{subequations}

One can observe several features of the second order solution. First, the angular dependencies of the first order were of $2\theta$ and $4\theta$. In the second order solution we get a new angular dependency, of $6 \theta$.
Another feature is related to the radial dependency of the solution. In the first order solution the different elements of the stress scales like $\frac{1}{r^2}$ or $\frac{1}{r^4}$, in the second order solution one can also find elements of $\frac{1}{r^6}$ and $\frac{1}{r^8}$.\\

We also get a correction to the Stress Concentration Factor (S.C.F): $K_t=3\left(1-\frac{5-2\nu+\nu^2}{4}\frac{\sigma_\infty}{Y}\right)$.\\

The main characteristic properties of the second order stress tensor are already reflected from the form \eqref{eq:Bstress} and the new singularities in $f_i$.
Explicit expressions for the stress tensor are derived in the attached Mathematica notebook.

\section{Nonlinear Crack}
\subsection{Mode I}\label{app:mode1}

The linear stress function of the Mode I crack is

\begin{subequations}
\begin{align}
    \phi^{\left(1\right)}\left(z\right)&=\frac{\sigma_\infty^{yy}}{4}\left(-z+2\sqrt{z^2-1}\right)+\frac{\sigma_{\infty}^{xx}}{4}z\\
        \psi^{\left(1\right)}\left(z\right)&=\frac{\sigma_{\infty}^{yy}}{2}\left(z-\frac{1}{\sqrt{z^2-1}}\right)-\frac{\sigma_{\infty}^{xx}}{2}z
\end{align}
\end{subequations}
where $\sigma_\infty^{xx}$ and $\sigma_\infty^{yy}$ are the stresses imposed at infinity.

We want to examine the contribution of the second order to the stress around the tip of the crack. Therefore we  expand the elements of the stress around the tip, i.e. around the point $\left(1+r \cos{\theta},r \sin{\theta}\right)$ in orders of $r$ and we get that asymptotic stresses take the form

\begin{subequations}
\begin{align}
    \sigma^{xx\left(1\right)}\left(r,\theta\right)&=\frac{\sigma_\infty^{yy}}{4\sqrt{2r}}\left[3\cos{\frac{\theta}{2}}+\cos{\frac{5\theta}{2}}\right]\\
    \sigma^{xy\left(1\right)}\left(r,\theta\right)&=\frac{\sigma_\infty^{yy}}{2\sqrt{2r}}\left[-\sin{\frac{\theta}{2}}+\sin{\frac{5\theta}{2}}\right]\\
    \sigma^{yy\left(1\right)}\left(r,\theta\right)&=\frac{\sigma_\infty^{yy}}{4\sqrt{2r}}\left[5\cos{\frac{\theta}{2}}-\cos{\frac{5\theta}{2}}\right]
\end{align}
\end{subequations}

As we showed before, the stresses at each order are affected by the stresses of lower orders. In the case of the mode I crack, the second-order stresses induced on the crack faces by the first-order stresses are

\begin{equation}
    \begin{split}
        \Delta\sigma^{yy\left(2\right)}\left(x\pm i\epsilon\right)=&\frac{\left(7-4\nu+\nu^2\right)\left[\left(\sigma^{xx}_{\infty}\right)^2-2\sigma^{xx}_{\infty}\sigma^{yy}_{\infty}-3\left(\sigma^{yy}_{\infty}\right)^2\right]}{16Y}\\
        i\Delta\sigma^{xy\left(2\right)}\left(x\pm i \epsilon\right)=&\pm\frac{\sigma^{yy}_{\infty} \left( \sigma^{xx}_{\infty}-\sigma^{yy}_{\infty}\right)}{8Y \left(x^2-1\right)^{3/2}}\left[6\left(\nu-1\right) x-\left(1+\nu\right)^2 x^3\right]
    \end{split}
\end{equation}\label{app:deltastress2}

The second-order stresses at infinity that are induced by the first-order stresses are
\begin{equation}
\begin{split}
    s^{xx\left(2\right)}=&\frac{-\left(\nu ^2-8 \nu +11\right) \left(\sigma_\infty^{ xx}\right)^2+\left(\nu ^2-4 \nu
   +7\right) \left(\sigma_\infty^{yy}\right)^2+4 \nu  (\nu +3) \sigma_\infty^{xx}\sigma_\infty^{yy}}{16 Y}\\
   s^{yy\left(2\right)}=&\frac{\left(\nu ^2-4 \nu +7\right) \left(\sigma_\infty^{ xx}\right)^2-\left(\nu ^2-8 \nu +11\right)
   \left(\sigma_\infty^{ yy}\right)^2+4 \nu  (\nu +3) \sigma_\infty^{xx}\sigma_\infty^{yy}}{16 Y}
\end{split}
\end{equation}\label{app:deltastress2inf}

Here we show the asymptotic stresses for $\nu=0$. Explicit derivation and expressions of the second-order correction to the stress field is given in \cite{Math}.

The asymptotic stress field takes the form

\begin{subequations}
\begin{align}
\sigma^{xx\left(2\right)}\left(r,\theta\right)&=\frac{\sigma^{yy}_\infty}{ Y}\left[\frac{g^{I}_1\left(\theta\right)}{\sqrt{r}}+\frac{g^{I}_2\left(\theta\right)}{r} +\frac{g^{I}_3\left(\theta\right)}{r^{3/2}}\right]\\
\sigma^{xy\left(2\right)}\left(r,\theta\right)&=\frac{\sigma^{yy}_\infty}{ Y}\left[\frac{g^{I}_4\left(\theta\right)}{\sqrt{r}}+\frac{g^{I}_5\left(\theta\right)}{r} +\frac{g^{I}_6\left(\theta\right)}{r^{3/2}}\right]\\
\sigma^{yy\left(2\right)}\left(r,\theta\right)&=\frac{\sigma^{yy}_\infty}{ Y}\left[\frac{g^{I}_7\left(\theta\right)}{\sqrt{r}}+\frac{g^{I}_8\left(\theta\right)}{r} +\frac{g^{I}_9\left(\theta\right)}{r^{3/2}}\right]\\
\end{align}
\end{subequations}

where $r=0$ denotes the crack tip and

\begin{subequations}
\begin{align}
    g^{I}_1\left(\theta\right)&=\frac{9 \cos \frac{\theta }{2} \left(61 \sigma ^{yy}_{\infty}-125 \sigma
   ^{xx}_{\infty}\right)+\cos \frac{5 \theta }{2} \left(95 \sigma
   ^{yy}_{\infty}-287 \sigma ^{xx}_{\infty}\right)+72 \cos\frac{9 \theta }{2}
   \left(\sigma ^{xx}_{\infty}-\sigma ^{yy}_{\infty}\right)}{512 \sqrt{2}}\\
   g^{I}_2\left(\theta\right)&=-\frac{1}{128} \left(96 \cos \theta -32 \cos 2 \theta +64 \cos 3 \theta +9 \cos 4
   \theta -57\right) \sigma ^{yy}_{\infty}\\
   g^{I}_3\left(\theta\right)&=\frac{15 \left(\cos \frac{3 \theta }{2}+3 \cos \frac{7 \theta
   }{2}\right) \left(\sigma ^{yy}_{\infty}-\sigma ^{xx}_{\infty}\right)}{128 \sqrt{2}}\\
    g^{I}_4\left(\theta\right)&=\frac{3 \sin \theta  \left[\cos \frac{3 \theta }{2} \left(29 \sigma
   ^{yy}_{\infty}-93 \sigma ^{xx}_{\infty}\right)+24 \cos \frac{7 \theta }{2}
   \left(\sigma ^{xx}_{\infty}-\sigma ^{yy}_{\infty}\right)\right]}{256 \sqrt{2}}\\
    g^{I}_5\left(\theta\right)&=-\frac{\left(-42 \sin 2 \theta+64 \sin 3 \theta +9 \sin 4 \theta \right) \sigma
   ^{yy}_{\infty}}{128}\\
      g^{I}_6\left(\theta\right)&=\frac{45 \sin \theta  \cos \left(\frac{5 \theta }{2}\right) \left(\sigma
   ^{yy}_{\infty}-\sigma ^{xx}_{\infty}\right)}{64 \sqrt{2}}\\
       g^{I}_7\left(\theta\right)&=\frac{-\cos \frac{\theta }{2} \left(403 \sigma ^{xx}_{\infty}+557
   \sigma ^{yy}_{\infty}\right)+\cos\frac{5 \theta }{2} \left(415
   \sigma ^{xx}_{\infty}-223 \sigma ^{yy}_{\infty}\right)-72 \cos \frac{9 \theta
   }{2} \left(\sigma ^{xx}_{\infty}-\sigma ^{yy}_{\infty}\right)}{512 \sqrt{2}}\\
        g^{I}_8\left(\theta\right)&=\frac{\cos ^2\left(\frac{\theta }{2}\right) \left(-153 \cos \theta +46 \cos 2 \theta +9
   \cos 3 \theta +82\right) \sigma ^{yy}_{\infty}}{32}\\
         g^{I}_9\left(\theta\right)&=\frac{15 \left(7 \cos \frac{3 \theta }{2}-3 \cos \frac{7 \theta
   }{2}\right) \left(\sigma ^{yy}_{\infty}-\sigma ^{xx}_{\infty}\right)}{128
   \sqrt{2}}
\end{align}
\end{subequations}

\subsection{Mode II}\label{app:mode2}
The linear stress function of the Mode II crack is

\begin{subequations}
\begin{align}
    \phi^{\left(1\right)}\left(z\right)&=-\frac{i\sigma_{\infty}}{2}\sqrt{z^2-1}\\
        \psi^{\left(1\right)}\left(z\right)&=\frac{i\sigma_{\infty}}{2}\left(\frac{2\; z^2-1}{\sqrt{z^2-1}}\right)
\end{align}
\end{subequations}

which gives the following asymptotic behaviour around the crack tip:
\begin{subequations}
\begin{align}
    \sigma^{xx\left(1\right)}\left(r,\theta\right)&=-\frac{\sigma_{\infty}}{4\sqrt{2r}}\left[7 \sin \frac{\theta }{2}+\sin \frac{5 \theta }{2}\right]\\
    \sigma^{xy\left(1\right)}\left(r,\theta\right)&=\frac{\sigma_{\infty}}{2\sqrt{2r}}\left[3 \cos \frac{\theta }{2}+\cos \frac{5 \theta }{2}\right]\\
    \sigma^{yy\left(1\right)}\left(r,\theta\right)&=\frac{\sigma_{\infty}}{4\sqrt{2r}}\left[- \sin \frac{\theta }{2}+\sin \frac{5 \theta }{2}\right]
\end{align}
\end{subequations}

Analogous to Mode I, the second-order stresses induced on the crack faces and at infinity by the first-order stresses are
\begin{subequations}
\begin{gather}
    \Delta\sigma^{yy\left(2\right)}\left(x\pm i\epsilon\right)=-\frac{\sigma^2_{\infty}}{4Y}\left(7-4\nu+\nu^2\right)\\
    i\Delta\sigma^{xy\left(2\right)}\left(x\pm i\epsilon\right)=0\\
    s^{xx\left(2\right)}=-\frac{\sigma^2_{\infty}}{4Y}\left(-5+8\nu+\nu^2\right)\\
    s^{yy\left(2\right)}=\frac{\sigma^2_{\infty}}{4Y}\left(7-4\nu+\nu^2\right)
\end{gather}
\end{subequations}

The second order correction to the stress function of Mode II crack is given by
\begin{subequations}
\begin{gather}
    \Omega^{\left(2\right)}\left(z\right)=\frac{1}{2}\left[\frac{L^{\left(2\right)}\left(z\right)}{\sqrt{z^2-1}}+M^{\left(2\right)}\left(z\right)-\Delta\sigma^{yy\left(2\right)}\right]\\
    \overline{\Phi^{\left(2\right)}}\left(z\right)=\frac{1}{2}\left[\frac{L^{\left(2\right)}\left(z\right)}{\sqrt{z^2-1}}-M^{\left(2\right)}\left(z\right)-\Delta\sigma^{yy\left(2\right)}\right]
\end{gather}
\end{subequations}
where $\Omega^{\left(2\right)}$ and $\Phi^{\left(2\right)}$ are defined by $\Psi=\psi'$, $\Phi=\phi'$, and $\Omega=\Phi+\Psi+z\Phi'$. $L$ and $M$ are determined by the boundary conditions at infinity and by demanding the continuity of the displacement outside of the crack
\begin{subequations}
\begin{gather}
    L^{\left(2\right)}\left(z\right)=\left(\Delta\sigma^{yy\left(2\right)}-s^{yy\left(2\right)}\right)z\\
    M^{\left(2\right)}\left(z\right)=\frac{1}{2}\left(s^{xx\left(2\right)}-s^{yy\left(2\right)}\right)
\end{gather}
\end{subequations}

Here we show the asymptotic stresses for $\nu=0$. Explicit derivation and expressions of the second-order correction to the stress field is given in \cite{Math}.

The asymptotic stress field takes the form

\begin{subequations}
\begin{align}
\sigma^{xx\left(2\right)}\left(r,\theta\right)&=\frac{\left(\sigma^{yy}_\infty\right)^2}{ Y}\left[\frac{g^{II}_1\left(\theta\right)}{\sqrt{r}}+\frac{g^{II}_2\left(\theta\right)}{r}\right]\\
\sigma^{xy\left(2\right)}\left(r,\theta\right)&=\frac{\left(\sigma^{yy}_\infty\right)^2}{ Y}\left[\frac{g^{II}_3\left(\theta\right)}{\sqrt{r}}+\frac{g^{II}_4\left(\theta\right)}{r}\right]\\
\sigma^{yy\left(2\right)}\left(r,\theta\right)&=\frac{\left(\sigma^{yy}_\infty\right)^2}{ Y}\left[\frac{g^{II}_5\left(\theta\right)}{\sqrt{r}}+\frac{g^{II}_6\left(\theta\right)}{r}\right]\\
\end{align}
\end{subequations}

where $r=0$ denotes the crack tip and

\begin{subequations}
\begin{align}
    g^{II}_1\left(\theta\right)&=-\frac{\left(5+52 \pi\right ) \left(3 \cos \frac{\theta }{2}+\cos \frac{5 \theta
   }{2}\right)}{64 \sqrt{2} \pi }\\
    g^{II}_2\left(\theta\right)&=\frac{1}{128} (96 \cos \theta +96 \cos 2 \theta +64 \cos 3 \theta +27 \cos 4
   \theta -43)\\
    g^{II}_3\left(\theta\right)&=-\frac{\left(5+52 \pi \right) \sin \theta  \cos \frac{3 \theta }{2}}{32 \sqrt{2} \pi
   }\\
    g^{II}_4\left(\theta\right)&=\frac{1}{128} \left(66 \sin 2 \theta +64 \sin 3 \theta +27 \sin 4 \theta \right)\\
    g^{II}_5\left(\theta\right)&=\frac{\left(5+52 \pi \right) \cos ^3\left(\frac{\theta }{2}\right)\left (2 \cos \theta -3\right)}{16 \sqrt{2}
   \pi }\\
    g^{II}_6\left(\theta\right)&=\frac{1}{32} \cos ^2\left(\frac{\theta }{2}\right) \left(11 \cos \theta -10 \cos 2 \theta
   -27 \cos 3 \theta +42\right)
\end{align}
\end{subequations}

\bibliographystyle{unsrt}
\bibliography{bib}

\begin{thebibliography}{10}

\bibitem{seung1988defects}
Hyunjune~Sebastian Seung and David~R Nelson.
\newblock Defects in flexible membranes with crystalline order.
\newblock {\em Physical Review A}, 38(2):1005, 1988.

\bibitem{broberg1999cracks}
K~Bertram Broberg.
\newblock {\em Cracks and fracture}.
\newblock Elsevier, 1999.

\bibitem{lidmar2003virus}
Jack Lidmar, Leonid Mirny, and David~R Nelson.
\newblock Virus shapes and buckling transitions in spherical shells.
\newblock {\em Physical Review E}, 68(5):051910, 2003.

\bibitem{armon2011geometry}
Shahaf Armon, Efi Efrati, Raz Kupferman, and Eran Sharon.
\newblock Geometry and mechanics in the opening of chiral seed pods.
\newblock {\em Science}, 333(6050):1726--1730, 2011.

\bibitem{mirabet2011role}
Vincent Mirabet, Pradeep Das, Arezki Boudaoud, and Olivier Hamant.
\newblock The role of mechanical forces in plant morphogenesis.
\newblock {\em Annual review of plant biology}, 62:365--385, 2011.

\bibitem{landau1959course}
Lev~Davidovich Landau and Eugin~M Lifshitz.
\newblock {\em Course of Theoretical Physics Vol 7: Theory of Elasticity}.
\newblock Pergamon press, 1959.

\bibitem{airy1862strains}
George~Biddell Airy.
\newblock On the strains in the interior of beams.
\newblock {\em Philosophical transactions of the Royal society of London},
  153:49--80, 1862.

\bibitem{Kupferman2021doubleI}
Raz Kupferman and Roee Leder.
\newblock Double forms: Part i. regular elliptic bilaplacian operators.
\newblock {\em arXiv preprint arXiv:2103.16823}, 2021.

\bibitem{Kupferman2021doubleII}
Raz Kupferman and Roee Leder.
\newblock Double forms: Part ii. on saint-venant compatibility and stress
  potentials in manifolds with boundary and constant sectional curvature.
\newblock {\em arXiv preprint arXiv:2104.05794}, 2021.

\bibitem{Nelson2002defects}
David~R Nelson.
\newblock {\em Defects and geometry in condensed matter physics}.
\newblock Cambridge University Press, 2002.

\bibitem{Kosmrlj2013mechanical}
Andrej Ko{\v{s}}mrlj and David~R Nelson.
\newblock Mechanical properties of warped membranes.
\newblock {\em Physical Review E}, 88(1):012136, 2013.

\bibitem{Bausch2003grain}
AR~Bausch, Mark~John Bowick, A~Cacciuto, AD~Dinsmore, MF~Hsu, DR~Nelson,
  MG~Nikolaides, A~Travesset, and DA~Weitz.
\newblock Grain boundary scars and spherical crystallography.
\newblock {\em Science}, 299(5613):1716--1718, 2003.

\bibitem{Tobasco2021curvature}
Ian Tobasco.
\newblock Curvature-driven wrinkling of thin elastic shells.
\newblock {\em Archive for Rational Mechanics and Analysis}, 239(3):1211--1325,
  2021.

\bibitem{Henkes2009statistical}
Silke Henkes and Bulbul Chakraborty.
\newblock Statistical mechanics framework for static granular matter.
\newblock {\em Physical Review E}, 79(6):061301, 2009.

\bibitem{muskhelishvili1946singular}
Nikola{\u\i}~Ivanovich Muskhelishvili and Jens Rainer~Maria Radok.
\newblock {\em Singular integral equations: boundary problems of function
  theory and their application to mathematical physics}.
\newblock Noordhoff, 1946.

\bibitem{muskhelishvili_problems}
NI~Muskhelishvili.
\newblock {\em Some basic problems of the mathematical theory of elasticity}.
\newblock Noordhoff, Groningen, 1953.

\bibitem{livne2008breakdown}
Ariel Livne, Eran Bouchbinder, and Jay Fineberg.
\newblock Breakdown of linear elastic fracture mechanics near the tip of a
  rapid crack.
\newblock {\em Physical review letters}, 101(26):264301, 2008.

\bibitem{bouchbinder2008weakly}
Eran Bouchbinder, Ariel Livne, and Jay Fineberg.
\newblock Weakly nonlinear theory of dynamic fracture.
\newblock {\em Physical Review Letters}, 101(26):264302, 2008.

\bibitem{bouchbinder2010weakly}
Eran Bouchbinder, Ariel Livne, and Jay Fineberg.
\newblock Weakly nonlinear fracture mechanics: experiments and theory.
\newblock {\em International journal of fracture}, 162(1-2):3--20, 2010.

\bibitem{moshe2014plane}
Michael Moshe, Eran Sharon, and Raz Kupferman.
\newblock The plane stress state of residually stressed bodies: A stress
  function approach.
\newblock {\em arXiv preprint arXiv:1409.6594}, 2014.

\bibitem{moshe2015elastic}
Michael Moshe, Eran Sharon, and Raz Kupferman.
\newblock Elastic interactions between two-dimensional geometric defects.
\newblock {\em Physical Review E}, 92(6):062403, 2015.

\bibitem{Bar-Sinai2020}
Yohai Bar-Sinai, Gabriele Librandi, Katia Bertoldi, and Michael Moshe.
\newblock Geometric charges and nonlinear elasticity of two-dimensional elastic
  metamaterials.
\newblock {\em Proceedings of the National Academy of Sciences},
  117(19):10195--10202, 2020.

\bibitem{gurtin1963generalization}
Morton~E Gurtin.
\newblock A generalization of the beltrami stress functions in continuum
  mechanics.
\newblock Technical report, BROWN UNIV PROVIDENCE RI DIV OF APPLIED
  MATHEMATICS, 1963.

\bibitem{thorne1973gravitation}
Kip~S Thorne, Charles~W Misner, and John~Archibald Wheeler.
\newblock {\em Gravitation}.
\newblock Freeman San Francisco, CA, 1973.

\bibitem{kolosov1909application}
GV~Kolosov.
\newblock On an application of complex function theory to a plane problem of
  the mathematical theory of elasticity.
\newblock {\em Yuriev, Russia}, 1909.

\bibitem{afek2005void}
Itai Afek, Eran Bouchbinder, Eytan Katzav, Joachim Mathiesen, and Itamar
  Procaccia.
\newblock Void formation and roughening in slow fracture.
\newblock {\em Physical Review E}, 71(6):066127, 2005.

\bibitem{ciarlet2005introduction}
Philippe~G Ciarlet.
\newblock An introduction to differential geometry with applications to
  elasticity.
\newblock {\em Journal of Elasticity}, 78(1):1--215, 2005.

\bibitem{koiter1966nonlinear}
Warner~Tjardus Koiter.
\newblock On the nonlinear theory of thin elastic shells.
\newblock {\em Proc. Koninkl. Ned. Akad. van Wetenschappen, Series B},
  69:1--54, 1966.

\bibitem{efrati2009elastic}
Efi Efrati, Eran Sharon, and Raz Kupferman.
\newblock Elastic theory of unconstrained non-euclidean plates.
\newblock {\em Journal of the Mechanics and Physics of Solids}, 57(4):762--775,
  2009.

\bibitem{kupferman2015metric}
Raz Kupferman, Michael Moshe, and Jake~P Solomon.
\newblock Metric description of singular defects in isotropic materials.
\newblock {\em Archive for Rational Mechanics and Analysis}, 216(3):1009--1047,
  2015.

\bibitem{docarmo2016differential}
Manfredo~P Do~Carmo.
\newblock {\em Differential Geometry of Curves and Surfaces: Revised and
  Updated Second Edition}.
\newblock Courier Dover Publications, 2016.

\bibitem{kirsch1898theorie}
C~Kirsch.
\newblock Die theorie der elastizitat und die bedurfnisse der festigkeitslehre.
\newblock {\em Zeitschrift des Vereines Deutscher Ingenieure}, 42:797--807,
  1898.

\bibitem{Math}
The full derivation is given in an attached mathematica notebook.

\bibitem{irwin1957analysis}
George~R Irwin.
\newblock Analysis of stresses and strains near the end of a crack transversing
  a plate.
\newblock {\em Trans. ASME, Ser. E, J. Appl. Mech.}, 24:361--364, 1957.

\bibitem{gol1974brittle}
Robert~V Gol'dstein and Rafael~L Salganik.
\newblock Brittle fracture of solids with arbitrary cracks.
\newblock {\em International journal of Fracture}, 10(4):507--523, 1974.

\bibitem{DUGDALE1960100}
D.S. Dugdale.
\newblock Yielding of steel sheets containing slits.
\newblock {\em Journal of the Mechanics and Physics of Solids}, 8(2):100--104,
  1960.

\bibitem{bouchbinder20091}
Eran Bouchbinder, Ariel Livne, and Jay Fineberg.
\newblock The 1/r singularity in weakly nonlinear fracture mechanics.
\newblock {\em Journal of the Mechanics and Physics of Solids},
  57(9):1568--1577, 2009.

\bibitem{bouchbinder2010autonomy}
Eran Bouchbinder.
\newblock Autonomy and singularity in dynamic fracture.
\newblock {\em Physical Review E}, 82(1):015101, 2010.

\bibitem{livne2010near}
Ariel Livne, Eran Bouchbinder, Ilya Svetlizky, and Jay Fineberg.
\newblock The near-tip fields of fast cracks.
\newblock {\em Science}, 327(5971):1359--1363, 2010.

\end{thebibliography}


\end{document}